# Electrical Characterization of High-k (k>115) Crystalline SrTiO$_3$ (STO) thin film integration with GaN with Nanomembrane Transfer Process

Md Tahmidul Alam[1], Kyoungjun Lee[2], Guangying Wang[1], Chang-beom Eom[2, a)] and Chirag Gupta[1, a)]

[1]Department of Electrical and Computer Engineering, University of Wisconsin-Madison, WI 53706, USA
[2]Department of Materials Science and Engineering, University of Wisconsin-Madison, WI 53706, USA

High-k (k>115), crystalline SrTiO$_3$ (STO) thin film was transferred on GaN for potential applications in power devices (transistor and diodes) by nanomembrane transfer method and the detailed electrical properties such as leakage current, C-V profiles, dielectric constant, frequency dispersion was reported from fabricated MOSCAP structures. The leakage current was negligible (under noise-level of tool) up to 6 V and 11 V for 50 nm and 200 nm STO membrane respectively A high-quality dielectric was indicated by the C-V profile, which showed almost negligible frequency dispersion in the frequency range of 10 kHz to 500 kHz. The dielectric constant was 50~82 with the 50 nm thick STO membrane and 115~186 in the 200 nm thick STO membrane. Thermal annealing of the membrane in ambient conditions at 250 °C for 2 hours led to a slight improvement in the dielectric constant (8–20%), albeit at the expense of degraded leakage current performance, as indicated by a reduction of 1 V ~ 3 V in the "no leakage region" of the IV curves after annealing. The possible physical mechanisms responsible for these changes were also analyzed and discussed.

Both lateral and vertical GaN power devices have been widely commercialized for power electronic applications due to GaN's high critical field (3.3 MV/cm), wide bandgap (3.4 eV) and high electron mobility[1–7]. However, the voltage blocking capability of these devices is still well behind the theoretical limits due to poor electric field management inside the semiconductor. The integration of a good-quality high-k dielectric (k>100) with GaN can significantly reduce the electric field stress hence can improve the voltage blocking capability[8–11]. Moreover, thicker high-k dielectrics can be used as gate dielectrics (compared to low-k oxides) without compensation in gate modulation, thus gate leakage and subthreshold swing (SS) can be reduced, on/off ratio can be increased, and it facilitates device scaling and ultra-large-scale-integration (ULSI)[12–17]. Earlier attempts to integrate high-k dielectrics with GaN include- sputtering of BaTiO$_3$ (BTO) (k =~ 280) on 10nm Al$_2$O$_3$/GaN (as field dielectric), atomic layer deposition (ALD) of HfO$_2$ (k =~25) (as gate dielectric), sputtering or e-beam evaporation of SrTiO$_3$ (STO) (k ~300) film on GaN to make low-leakage Schottky diodes[18–22]. However, these high-energy direct depositions (k>25) produce polycrystalline films and damage the GaN interface. To protect the GaN interface from damage, a thin (10nm) Al$_2$O$_3$ (low-k, k~10) layer is used that lowers the effective dielectric constant. Moreover, polycrystallinity in perovskite dielectrics (MTiO$_3$, M=Ba, Sr) leads to lower dielectric constant (k) compared to their crystalline counterparts, also introduces traps, defects, dislocations and reduces reliability[23,24]. The nanomembrane transfer process or remote epitaxy is a promising alternative to integrate high-k crystalline nanomembranes with GaN with minimal damage to the interface[25,26]. Recently, J. Ji et al. integrated crystalline STO on AlGaN/GaN HEMT by nanomembrane transfer method and utilized it as gate dielectric, achieved near-theoretical subthreshold swing (62 mV/dec) and high on-off ratio (2.6 ×10$^7$)[27]. However, the detailed electrical characterization of crystalline STO nanomembrane-on-GaN such as dielectric constant (k), C-V (capacitance-voltage) characteristics, leakage current, frequency dispersion etc. are not available in literature. In this work, we integrated high-quality, high-k (k>115) crystalline STO nanomembrane (50 nm and 200 nm) on GaN by nanomembrane transfer method and reported the detailed electrical characteristics. In our fabricated STO membrane there was no leakage current (< noise-level of the tool) up to 6V and 11V for 50 nm and 200 nm membrane respectively, the dielectric constant was >50 (50 nm) and >115 (200 nm). There was no frequency dispersion of capacitance and dielectric constant between 10 kHz and 500kHz, the capacitance value was almost perfectly proportional to the area of the contact pad- indicating the high-quality of the dielectric. STO was chosen as the dielectric material because of its high bulk dielectric constant (k>300), optical transparency in the visible range (for potential applications in optical devices) and excellent chemical and thermal stability[28,29].

The growth and transfer process of the nanomembrane is depicted in Fig. 1. The membrane (50nm and 200nm) was grown on 10nm SCAO (Scandium Aluminum Oxide, ScAlO$_3$)/ bulk STO (001) substrate by pulsed laser deposition (PLD) at 750°C. The 10nm SCAO layer was almost lattice matched to STO and water soluble. Following membrane growth, platinum (Pt) electrodes were sputtered on the membrane as contact pads (pad diameter = 50 μm and 100 μm). Subsequently, the sample was immersed in deionized (DI) water, resulting in the separation of the membrane from the bulk material. The detached membrane then floated on the water surface. Thereafter, a 2 μm n+ GaN / 2 μm UID GaN/20 nm AlGaN/ sapphire substrate (PSS) sample was dipped in the water and was gradually raised above the surface so that floating membrane clung to its n+ GaN face. Following that, ohmic contacts to the n+ GaN were made. Then, the membrane was annealed in the atmosphere at 250 °C for two hours. The membrane was then electrically characterized, both before and after annealing.

After transfer, there were some cracks on some part of the membrane however there was enough "clean" surface as well (> 50%) to fabricate devices (Fig. 2). The surface of the STO membrane was found to be atomically smooth in AFM images (Fig. 2). XRD measurements showed the existence of single-crystalline STO (200) and STO (100) peaks as depicted in Fig. 3. The leakage current through the membrane was characterized using IV measurements in a B1505A SMU, as presented in Fig. 4.. The reverse leakage through the membrane was below noise-level up to -7 V and -14 V for 50 nm and 200 nm thick STO respectively before annealing. However, after annealing the leakage threshold reduced to -6 V and -11 V respectively. This is possibly because the annealing process improved the crystalline quality of the membrane, leaving some more vertical filaments that contributed to leakage paths.

Capacitance-voltage (C-V) measurements were performed with different pad sizes to determine the dielectric constant, frequency dispersion and quality of the dielectric. The voltage range of the C-V measurement was always restricted to the "no-leakage zone" of the membrane, i.e., between 0 V to -5 V to ensure the leakage current does not impact measured value of the capacitance. Fig. 5 shows that the capacitance and voltage profile of the membrane remain remarkably stable with frequencies ranging from 10 kHz to 500 kHz. The C-V profile with 50 μm and 100 μm pad sizes indicates that the capacitance is almost perfectly proportional to the pad area (Fig. 6). The absence of frequency dispersion and near-perfect scalability of capacitance with pad area implies the excellent dielectric quality of the STO membrane. Fig. 6 also depicts the approximate dielectric constant of the membrane extracted from the simple parallel plate formula (ignoring the depletion capacitance of GaN and the interface capacitance). It was found that the dielectric constant was at least 95 (200 nm membrane). The extracted value of k was almost equal for 50μm and 100μm pad size. It should be noted that a variable k was found with the applied bias, however in reality the k-value of the membrane should be constant. The apparent variation in the value of k with applied bias was attributed to the variable depletion width of GaN with different bias voltages, which was ignored in this calculation.

To extract a more accurate value of the dielectric constant, the depletion capacitance of GaN should be taken into account. Therefore, we introduced a simple analytic-numeric method to determine the value of the dielectric constant from C-V measurements. In this method, the value of k was swept and the value of GaN depletion width ($d_{GAN}$) was extracted by two different approaches. The intersection point of the two k vs $d_{GaN}$ curves plotted from two different methods was considered to contain the true dielectric constant. First, the measured capacitance ($C_M$) between the Pt electrodes and the ohmic pads was considered to be a series equivalent of the GaN depletion capacitance ($C_{GaN}$) and the STO capacitance ($C_{STO}$) as shown in Fig. 7. The parallel plate formula was used ($C = \frac{k\varepsilon_o A}{d}$) to determine the value of each individual capacitance-

$$\frac{1}{C_M} = \frac{t_{STO}}{k_{STO}\varepsilon_o A} + \frac{d_{GaN}}{k_{GaN}\varepsilon_o A}$$

Here, $t_{STO}$ is the thickness of the STO nanomembrane, A is the area of the pad electrode, $\varepsilon_o$ is the permittivity of free space, $C_M$ was taken at zero bias. $k_{GaN}$ and $k_{STO}$ are the dielectric constants of STO nanomembrane and GaN respectively. $k_{GaN}$ was assumed to be 10. The dielectric constant of STO ($k_{STO}$) was swept between 50-300 and the extracted depletion thickness of GaN ($d_{GaN}$) was plotted in Fig. 9. Second, the area under the electric field curve was equated with the built-in potential ($V_{bi}$) of the MOSCAP. The built-in potential was found from the x-intercept of the $\frac{1}{C_M^2}$ vs voltage curve as shown in Fig. 8(a) ($V_{bi}$ = ~2.4 V). The electric field profile inside the STO and GaN depletion layers was obtained by solving the Poisson's equation from the depletion charge density of GaN ($N_D^+ = 1 \times 10^{19}$ cm$^{-3}$) and the thickness of the STO layer, as shown in Fig. 8(b). The discontinuity in the electric field at the STO/GaN interface originates from the different dielectric constants of the materials and the boundary condition, $k_{STO}E_{STO} = k_{GaN}E_{GaN}$. The following equation was derived-

$$V_{bi} = t_{STO}\frac{eN_D d_{GaN}}{k_{STO}\varepsilon_o} + \frac{eN_D d_{GaN}^2}{2k_{GaN}\varepsilon_o}$$

Here, $e$ is the charge of electron. The dielectric constant of STO ($k_{STO}$) vs depletion thickness of GaN ($d_{GaN}$) curve obtained from this equation was overlayed with the $k_{STO}$- $d_{GaN}$ profile extracted from series capacitance method as describe earlier (Fig.9). The intersection point of these curves suggested a dielectric constant of 115~186 for 200nm STO nanomembrane in multiple measurements with different pad sizes. For 50nm STO nanomembrane the value was between 52~82.

The value of the dielectric constant showed dependency on the STO thickness and contact pad size. Thicker STO thickness (200 nm) and smaller pad size (50μm) resulted in higher k-value in all measurements. The thickness dependance is consistent with previous reports and arises possibly due to the presence of "dead layers" on both sides of the film, the GaN interface and the metal contact interface[30–32]. The dead layers consist of defects, dislocations, vacancies and have a low dielectric constant. As the thickness of STO membrane increases, the percentage contribution of the dead layers reduces therefore a higher dielectric constant is observed. On the other hand, the k-value stays in the lower range (115~140) in 100 μm contact pads compared to 50μm pads (150~186) in the 200 nm membrane. The 50 nm nanomembrane had similar trends as well, the range of k was 50~60 in 100 μm contact pads whereas it was 70~82 in 50 μm pads. The origin of this trend is not clearly understood however it may be hypothesized that, bigger contact pads might have damaged the STO membrane a bit more compared to smaller ones during metal deposition or nanomembrane transfer process leading to lower crystallinity and degradation of the dielectric constant. One more trend was observed, the value of k was slightly higher after annealing, the k-value was 105~160 and 31~38 in 200 nm and 50 nm STO membrane respectively before annealing. A possible reason for improved k after annealing was that the annealing process might have healed some damages of the membrane that occurred during metal deposition or transfer process.

Different annealing temperatures and ambient conditions might play a crucial role in the damage recovery process and facilitate more optimization of the membrane quality and the dielectric constant. Our future studies are anticipated to focus on this optimization and contain the relevant results.

In conclusion, 200 nm and 500 nm single-crystalline STO nanomembranes (confirmed by XRD measurements) were transferred on GaN for potential applications as gate-dielectric (transistors) and field dielectric (transistors and diodes). The detailed electrical properties of the membrane were reported such as leakage current, capacitance-voltage, frequency dispersion, dielectric constant and its dependance on membrane thickness, contact pad size and annealing. A simple analytic-numeric method was introduced to evaluate the dielectric constant of a thin film. The transferred nano-membrane had negligible leakage current up to 6 V and 11 V for 50 nm and 200 nm thick STO membrane respectively. There was almost no frequency dispersion in the C-V profile in the range of 10 kHz- 500 kHz, and the value of the capacitance scaled proportionally with contact area- indicating a high-quality dielectric. The measured dielectric constant in 50 nm and 200 nm STO membrane was 50~82 and 115~186 respectively. It was found that smaller pad size and higher thickness of the membrane favor higher dielectric constant and the possible physical reason behind this was explained.

We gratefully acknowledge the support by NSF (National Science Foundation) through MRSEC (Materials Research Science and Engineering Center), University of Wisconsin-Madison, program: DMR1720415

**FIGURES**

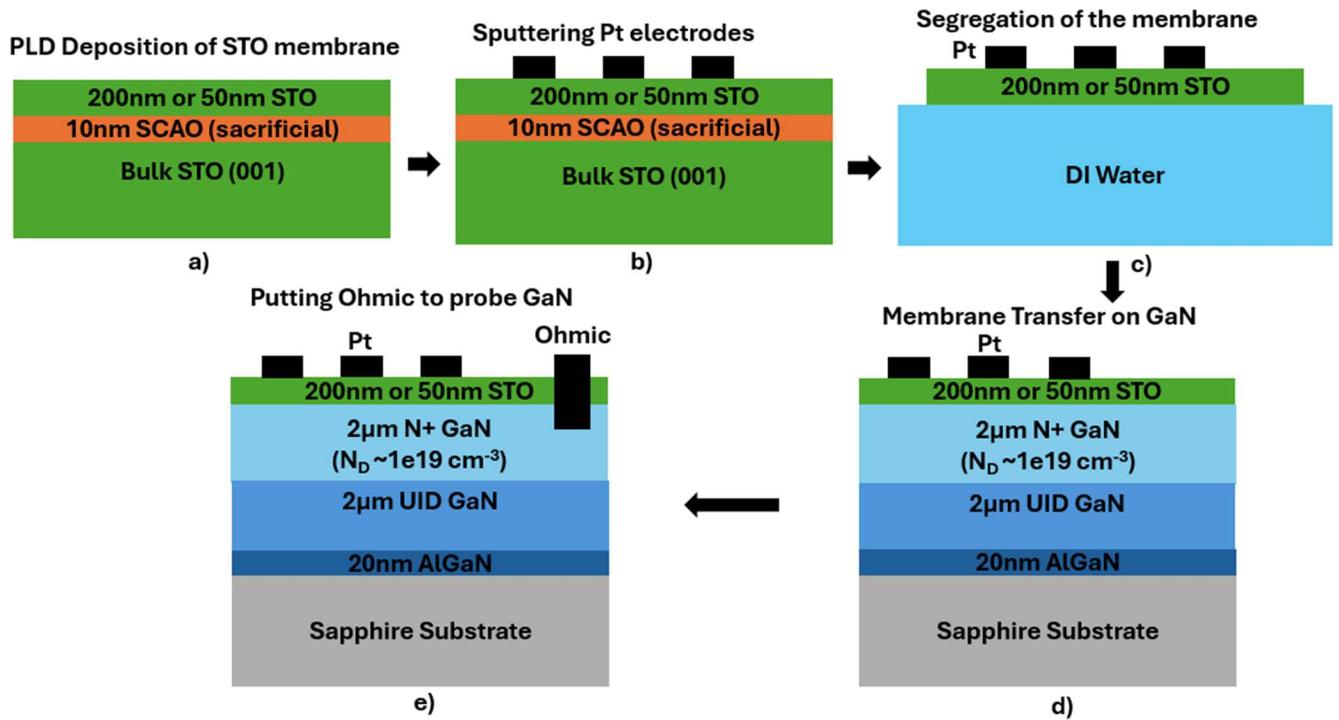

Fig. 1. The transfer process of the STO nanomembrane on GaN. a) Growth of the by pulsed laser deposition. b) Deposition of Pt electrodes by sputtering. c) Segregation of the membrane by dissolving the sacrificial layer. d) Transferring the membrane on GaN surface. e) Contacting the n+ GaN layer by an ohmic (Final structure).

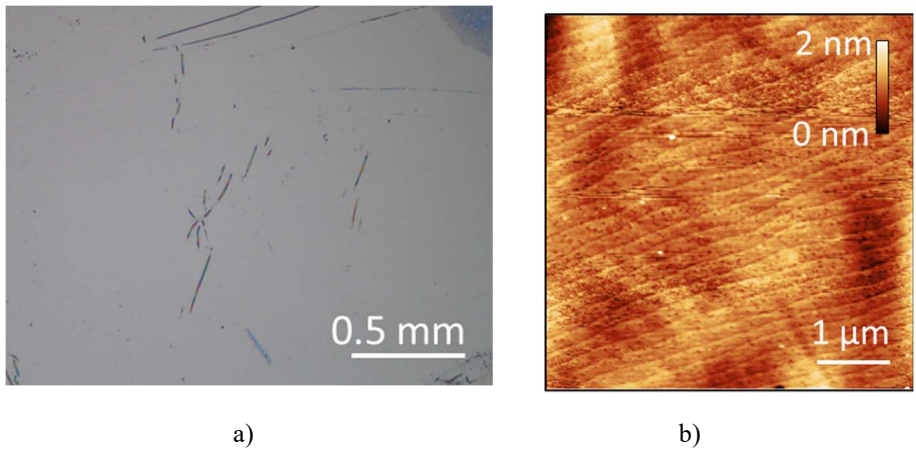

Fig.2. a) Surface of the transferred STO nanomembrane membrane on GaN (optical microscope). b) AFM image of the "clean area" showing atomically smooth surface.

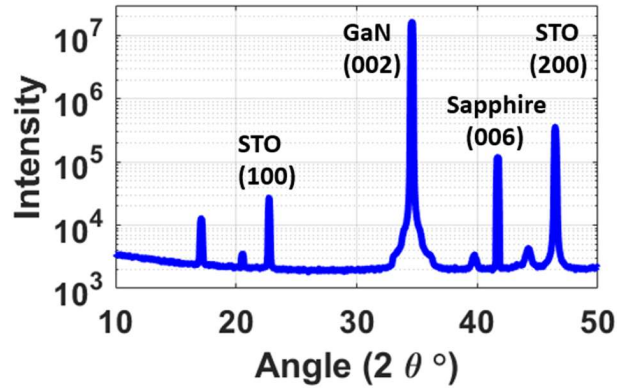

Fig.3. XRD measurements show the presence of crystalline STO (100) and STO (200) peaks.

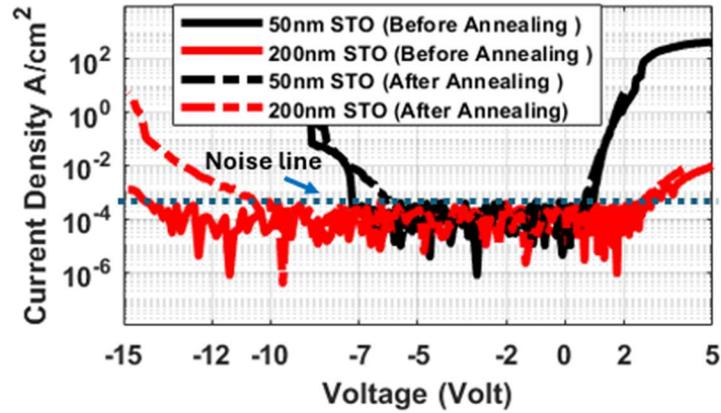

Fig. 4. Leakage current through the membrane. The thinner membrane had higher leakage. After annealing the membrane starts to leak at lower voltage.

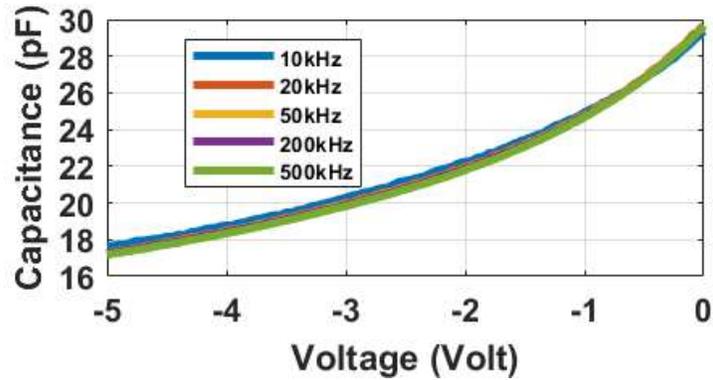

Fig. 5. C-V profile in the frequency range 10 kHz – 500 kHz ($t_{STO}$ = 200nm). Almost no frequency dispersion was observed.

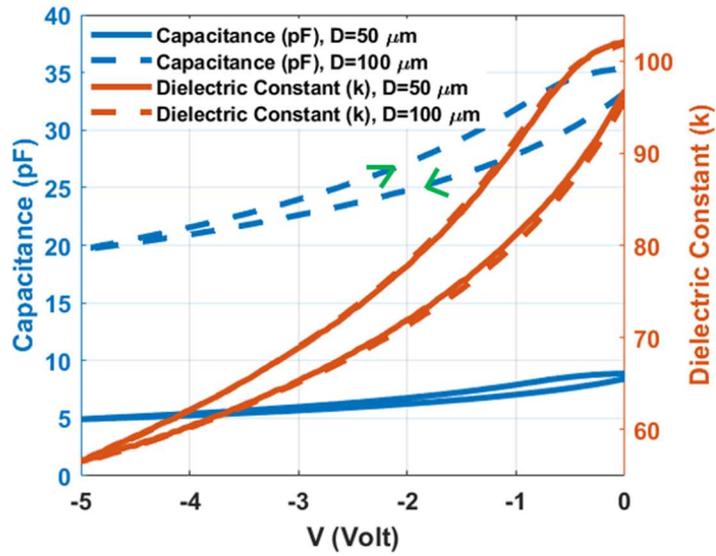

Fig. 6. Capacitance is near-perfectly proportional to contact pad area. Estimation of the dielectric constant from parallel plate formula ignoring the GaN depletion.

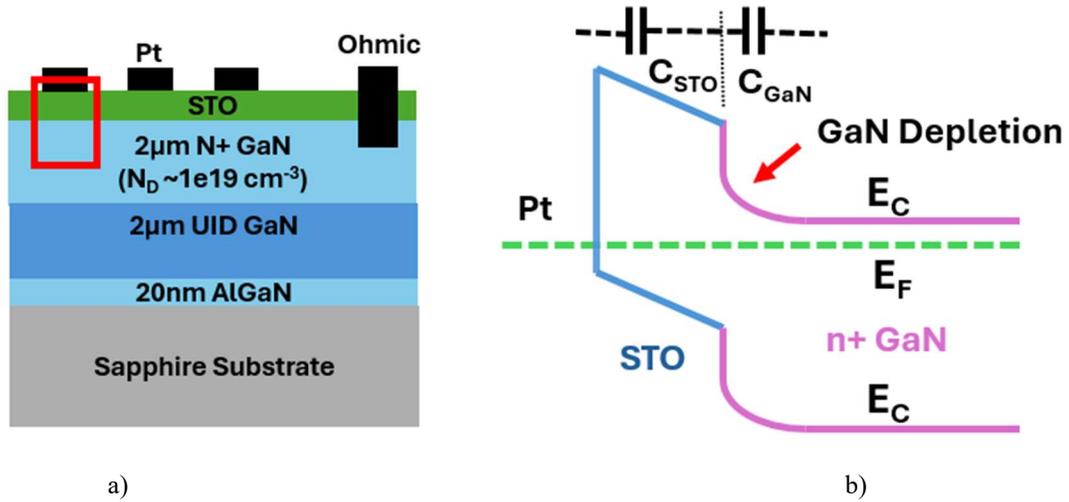

a)  b)

Fig. 7. a) The cross-section of the fabricated structure. b) The band diagram of the red box of the cross-section.

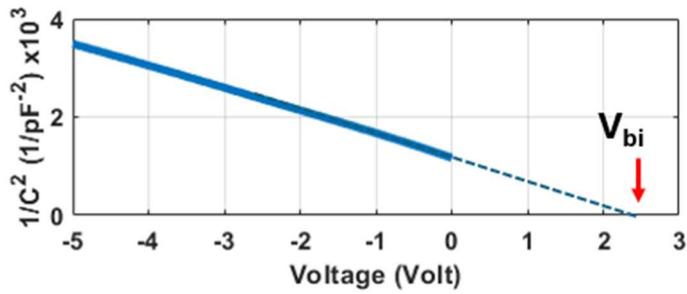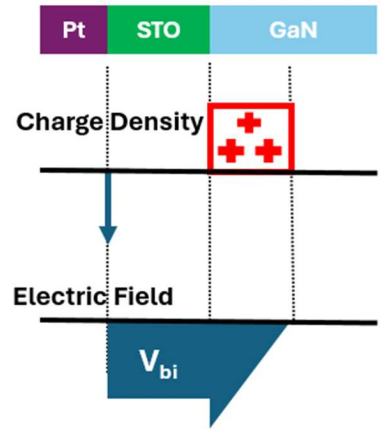

a)

b)

Fig. 8. a) Built-in potential ($V_{bi}$) extracted from the x-intercept of the $\frac{1}{C_M^2}$ vs voltage plot. b) Charge and electric field profile inside the MOSCAP structure. Built-in potential = area under the electric field.

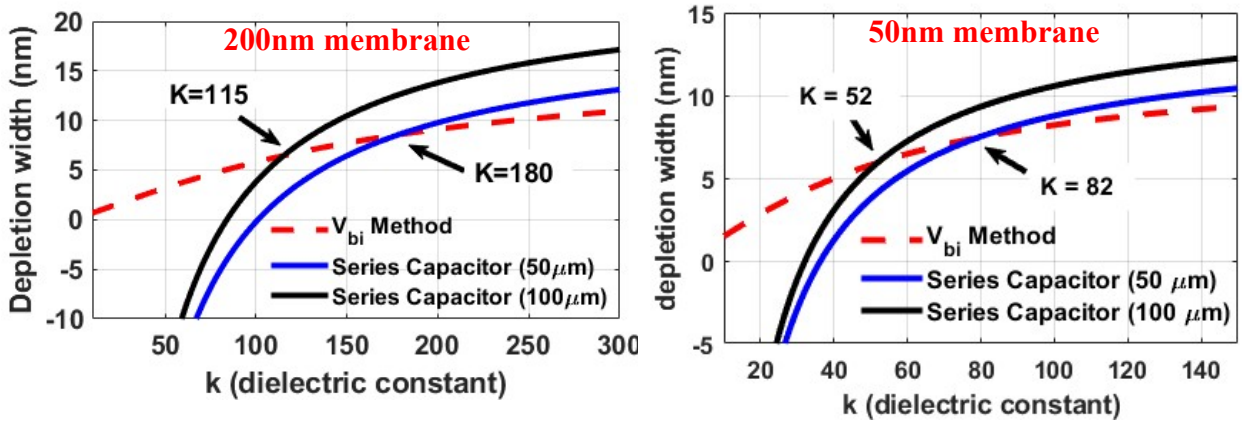

Fig. 9. Determination of k from the intersection of $d_{GaN}$ vs k plots found from built-in potential method and series capacitor method. Smaller contact pads were measured to have higher k. a) 200nm STO nanomembrane had k in 115-186. b) 50nm STO nanomembrane had k in 52-82.

11